\documentclass[preprint2]{aastex}
\usepackage{geometry}
\usepackage{natbib}
\begin{document}

\title{Temporal behavior of the SO 1.707 $\mu$m ro-vibronic emission band \\
in Io's atmosphere}
\shortauthors{Laver et al}
\author{Conor Laver, Imke de Pater}
\affil{Department of Astronomy, 601 Campbell Hall, University of California, \\ Berkeley, CA 94720}
\author{Henry Roe}
\affil{Division of Geological and Planetary Sciences, California Institute of Technology,\\ Pasadena, CA 91125}
\author{Darrell F. Strobel}
\affil{Department of Earth and Planetary Sciences and Department of Physics and Astronomy, Johns Hopkins University,\\ Baltimore, Maryland 21218}

\noindent
\\
\textbf{Abstract:}
We report observations of the ro-vibronic  $a^{1}\Delta\rightarrow X^{3}\Sigma^{-}$ transition of SO at 1.707 $\mu$m on Io. These data were taken while Io was eclipsed by Jupiter, on four nights between July 2000 and March 2003.  We analyze these results in conjunction with a previously published night to investigate the temporal behavior of these emissions. The observations were all conducted using the near-infrared spectrometer NIRSPEC on the W.M. Keck II telescope. The integrated emitted intensity for this band varies from 0.8 x $10^{27}$ to 2.4 x $10^{27}$ photons/sec, with a possible link to variations in Loki's infrared brightness. The band-shapes imply rotational temperatures of 550-1000K for the emitting gas, lending further evidence to a volcanic origin for sulfur monoxide. An attempt to detect the $B^{1}\Sigma\rightarrow X^{3}\Sigma^{-}$  transition of SO at 0.97 $\mu$m was unsuccessful; 
simultaneous  detection with the 1.707 $\mu$m band would permit determination of the SO 
column abundance.
\\

\section{Introduction}

Io, the most volcanic of Jupiter's moons, has been scrutinized from
both ground-based telescopes and spacecraft flybys for many years. In 2003, Galileo completed its extended mission to the Jovian system which provided a wealth of information, and an insight into the nature of Io's atmosphere and surface. Like most good science, the data raised many new questions, as it answered old ones. This volatile moon provides
us with an intriguing testbed for examining the effects of tidal stresses
on massive bodies, while its continuous volcanic activity (e.g. \cite{marchis05}, \cite{rathbun06}) results
in an ever-changing atmosphere \citep{lell96}. It remains our most easily observed
example of active volcanism in a low pressure environment, and of
a dynamically supported atmosphere; an analog which may be particularly
relevant to Martian and Lunar history.

The first detection of Io's atmosphere was made by the Pioneer 10
flyby which detected an ionosphere near the terminator \citep{kliore74}. The first atmospheric species, sulfur dioxide, was identified
in Voyager IRIS spectra by \citet{pearl79}, and confirmed
at millimeter and UV wavelengths by \cite{lell90}
and \cite{ball94} respectively. The observations
and subsequent analysis posed two major questions about the atmosphere
which are yet to be fully resolved. What are the relative contributions
to the atmosphere from volcanic eruptions, sublimation, and sputtering? And how localized is
the atmosphere?
It is widely agreed that all the above mechanisms must play a part although to what degree
is not yet well understood.  The presence of spatial variations was confirmed by
\cite{mcgrath00}, where they used spatially
resolved HST observations to confirm an atmospheric density enhancement
over Pele, a volcanic hotspot, compared to a quiescent location,''T3''. This local atmospheric enhancement near volcanic plumes has since been modeled in detail for a variety of plume types \citep{zhang03}.
Currently the exact properties and dominant processes of Io's global
atmosphere are still debated. Numerical models (e.g. \citealp{strobel94}, \citealp{wong96}, \citealp{wong00}), provide an estimate of the
vertical thermal and chemical profile; however, these models depend on the local surface temperature and pressure, which are not well constrained. 

While the SO$_{2}$ component of the atmosphere has been studied in
depth, identification of other gases in the atmosphere has been slow. The initial detection of SO came in the millimeter when \cite{lell96} identified it's rotational lines at 220 and 138GHz.
The first detection of ro-vibronic transitions of the sulfur monoxide radical
in the infrared at 1.707$\mu m$ was reported in \cite{depater2002} (henceforth referred to as DP02).

The importance of quantifying the levels of SO was emphasized
by \cite{summers96}. They suggested that SO's vapor pressure may be high enough that the gas may not condense out on Io's night side, which supported the predictions of 
\cite{ingersoll89}. It could, along with O$_{2}$, be
an important buffer gas in the atmosphere, allowing nanobars pressures
to exist following the collapse of the detected but condensable SO$_{2}$
atmosphere.  However SO is expected to be rapidly removed from the atmosphere through chemical reactions on the surface \citep{lell96}, and hence detections of this gas while Io is in eclipse imply that it must be constantly replenished. Volcanic production of SO via the quenching of the gases in high pressure vents \citep{zolotov98a} was shown to be able to support the observed ratios of 3-10\% by volume of SO/SO$_{2}$. Thus we should expect localized enhancements of SO abundance near such hotspots.

The SO ro-vibronic infrared band was detected in emission
in Io's near IR spectrum (1.707$\mu m$), while the satellite was in Jupiter's shadow. DP02 fit a rotational temperature of 1000K to
the band using lab spectra from \cite{setzer99} along
with line strengths from \cite{bellary87}, supporting \cite{zolotov98a}'s theory that the most likely source of this gas was a
high temperature vent. This current paper, a follow up to DP02, identifies this band on additional nights, supports their rotational temperature estimates and presents evidence of variability in the band emission.

\section{Observations}

\begin{figure}
\includegraphics[scale=0.7,bb= 80 20 400 380]{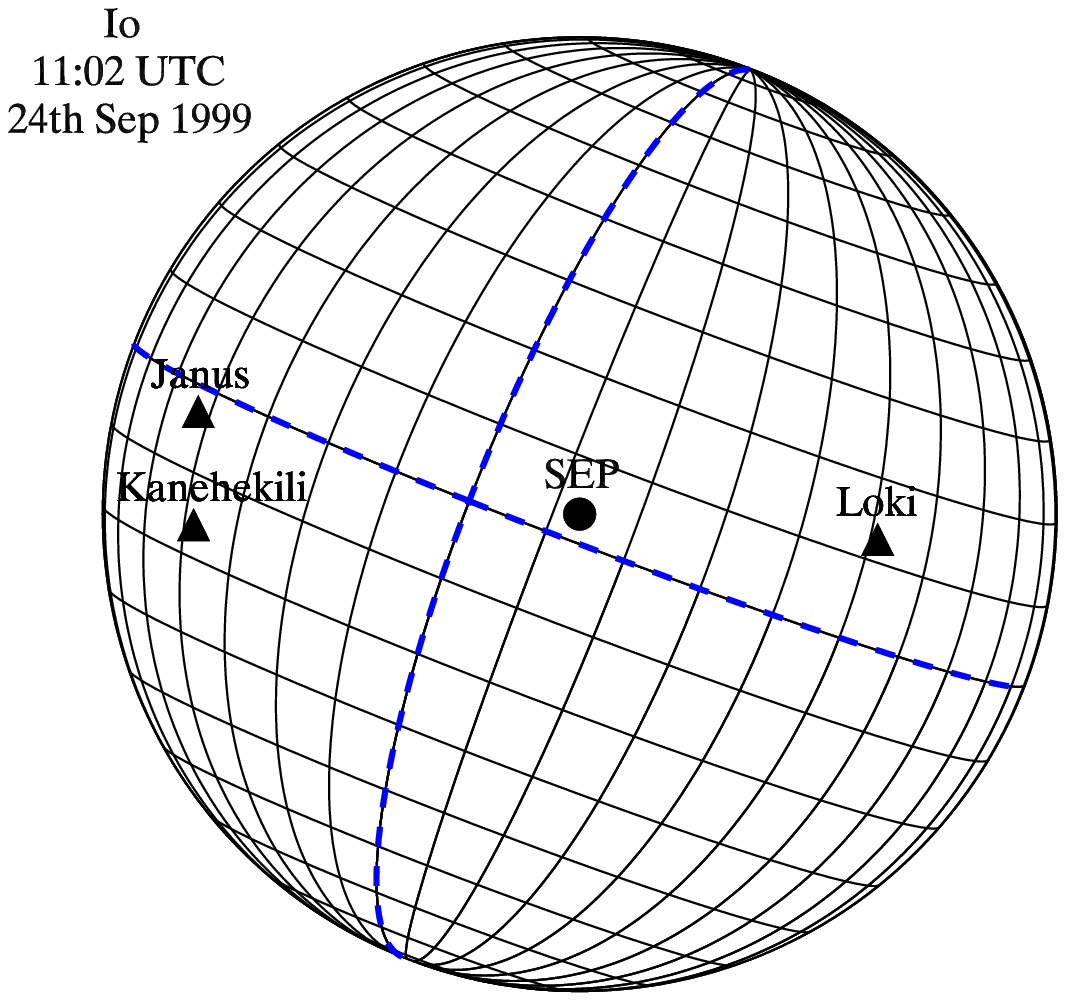}
\includegraphics[scale=0.7,bb= 80 20 425 380]{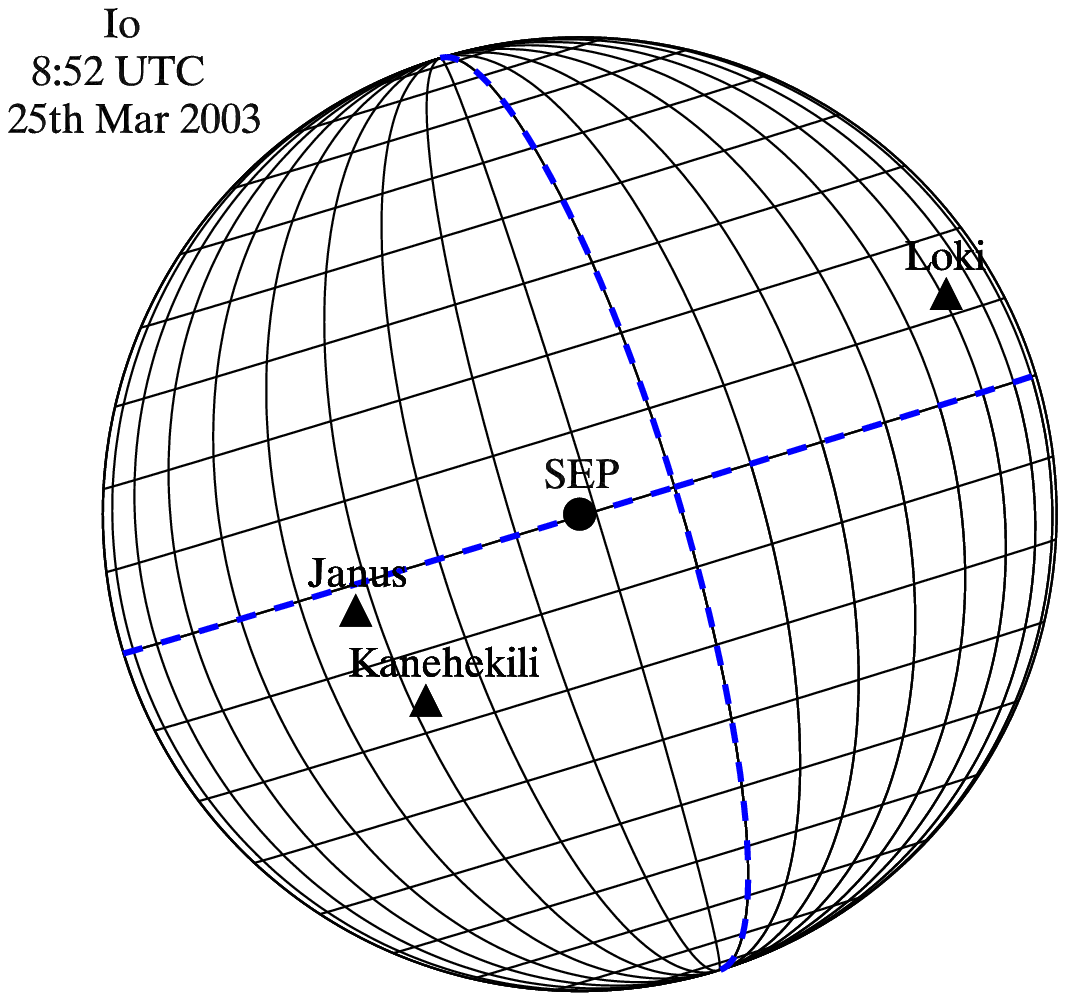}
\caption{\label{fig:coords}Observed face of Io during 1999 and 2003 eclipses. The dashed lines indicate 0 degrees longitude and latitude. Some of the most prominent hotspots and the sub-earth point (SEP) are indicated. The 2000 and 2002 eclipses present nearly identical faces as the 1999 data, as these are all eclipses in ingress, whereas the 2003 eclipse is in egress.}
\end{figure}

We present observations of Io taken while the satellite was in Jupiter's
shadow on four nights, from 2000 to 2003, and compare these to the
1999 data presented in de Pater et al 2002. The data were taken using
the near infrared spectrometer NIRSPEC on the W.M.Keck II telescope
at Mauna Kea, Hawaii, which is equipped with a 1024 x 1024 InSb ALADDIN
detector. The data presented here used the NIRSPEC-6 (N6: 1.558-2.315$\mu$m)
filter in low resolution echelle spectroscopy (R=2000) mode. The spectral region of the N6 filter analyzed for this paper was 1.6 - 2.0 $\mu$m. The total exposure
time in each case was a few minutes, which led to the detection
of the SO band at 1.707$\mu$m on all of the nights, although,
as we will show later, at various strengths. In all cases Io was observed
during the entire time it was in Jupiter's shadow, which was close to an hour in
duration on each night. The typical viewing geometry of Io is shown in Fig. \ref{fig:coords} , which shows some of the hotspots visible on Io going into and emerging from eclipse (2003 data). 

A list of observations and settings is given in Table \ref{tab:obs}.
We used a slit size of 0.76'' x 42'', the widest available, with 0.144'' pixels. Io
subtends $\ensuremath{\sim}$1'' and hence is larger than the slit. 
We also used a slit camera (SCAM) which has a 256 x 256 pixel HgCdTe
PICNIC array of 0.18'' pixels, which gives a field of view of 46''
x 46''. These images were used to confirm Io's position on the slit. One of the data sets was taken with the adaptive optics system, on 12th Nov 2002. These data are discussed in detail by \cite{depater07}. In this paper we report the disk integrated total flux density of these measurements.

\begin{deluxetable}{ccccccc}
\tablecaption{\label{tab:obs}Observations of Io in eclipse taken between 1999-2003}
\tablehead{\colhead{Date(UT)} & \colhead{Time(UT)} & \colhead{Object} & \colhead{Int Time(s)} & \colhead{SEL\tablenotemark{1}(deg)} &\colhead{Ang Size(arcsec)}& \colhead{Airmass}  } 
\startdata
9/24/1999 & 11:02 & Io in eclipse \tablenotemark{a} & 240 & 348 & 1.236 & 1.065   \\
9/24/1999 & 09:18 & SJ9182 \tablenotemark{b}& 60 & --- & --- &1.384\\
7/19/2000 & 14:18 & Io in eclipse & 150 & 347 & 0.91 & 1.95 \\
7/19/2000 & 12:55 & HD201941 & 75 & --- & --- &1.1 \\
11/12/2002 & 14:21 & Io in eclipse \tablenotemark{c} & 90 & 346 & 0.98 &1.13  \\
11/12/2002 & 12:53 & HIP44336 & 180 & --- & --- &1.38 \\
12/21/2002 & 12:44 & Io in eclipse & 360 & 348 & 1.098 &1.03 \\
12/21/2002 & 13:14 & HIP 45874 & 40 & --- & --- & 1.01  \\
3/25/2003 & 08:52 & Io in eclipse & 180 & 13 & 1.071 &1.13 \\
3/25/2003 & 08:40 & SAO 98024 & 120 & --- & --- &1.1 \\
\enddata
\tablenotetext{1}{Io's Geodetic Sub Earth Longitude}
\tablenotetext{a}{published in DP02}
\tablenotetext{b}{IR standard \cite{persson98}}
\tablenotetext{c}{Disk averaged AO data, see \cite{depater07} for full details}
\end{deluxetable}

All data were reduced normally: flat fielded, bad pixels removed, calibrated on neon and
argon lines, and  fine-tuned with telluric OH. The resulting calibration is correct to $<$1\AA. To remove telluric emission lines we subtracted 'nodded'  pairs of spectra which were taken in quick succession. 
We then averaged 15 pixels over Io (approximately twice the full width at half maximum) and subtracted the sky background, to obtain the highest signal-to-noise spectrum possible. We corrected for telluric absorption lines by dividing by either a normalized G or
A star spectrum. The spectra were then multiplied by an appropriate normalized stellar model to correct for the known stellar features.  G stars
are moderately flat in this region although they have many small features. Alternatively, A stars are smoother with only a few well known pronounced features which can be easily modeled. The models were also used for flux calibration and are discussed further below. All spectra were rectified and partially processed using routines from the REDSPEC pipeline developed for NIRSPEC reductions by the UCLA infrared lab, (see http://www2.keck.hawaii.edu/inst/nirspec/redspec/) Further analysis was performed using IDL routines written by the authors. 

The flux calibration of the spectra proved to be a challenging problem, due largely to the fact that the largest slit width for NIRSPEC
is 0.76'' compared to Io's angular size of $\sim$1''.
In DP02, this problem was tackled by calibrating Io's flux density in the SCAM (slit camera) images, by observing the G4 infrared standard star SJ9182 \citep{persson98}. Io's continuum level in the reduced spectra was then scaled to match the SCAM flux level, which is measured in the same filter bandpass and along the same optical path. The 2002 AO data from DP06 were also calibrated in this manner. The lack of suitable SCAM images taken on the other nights in 2000-2003 prevented a similar method being used.

In this paper we directly calibrated the stellar spectra using 2MASS. We used stellar models created by Kurucz (http://kurucz.harvard.edu/stars), which we reddened using the Fitzpatrick 1999 extinction curve, and flux calibrated to match the 2MASS magnitudes. The telluric calibration stars, despite not being published IR standards, all have been measured by the 2MASS survey, and all have the highest 'AAA' photometric stability rating. The photometric errors quoted for these stars are all $<$ 5\%, which is adequate as this is dwarfed by the uncertainties due to centering of the object on the slit and variations in atmospheric conditions. 

For the non-AO Io data we corrected for the overfilling of the slit by convolving a model disk of Io,  with a model gaussian PSF with a width equal to the average seeing on each night. Io's angular size, which varies with distance to earth, ranged from 0.91" to 1.24" on the various nights. The average seeing was 0.50",0.53",0.65"  and 0.70" for the 1999, 2000, 2002(Dec) and 2003 nights respectively. We could then estimate the percentage of light which entered the slit (69-80\%) and adjust the photometry appropriately.
The photometric errors after this correction are dominated by the variation of our calibration standards due to the atmosphere and positioning of the targets on the slit. We estimate the uncertainty to be $\sim$ 25-30\%. Obviously this is not ideal, however it is the best achievable given the data set, and is sufficient for our purposes. To ensure consistency we reanalyzed the 1999 data, using the flux calibration method described above. In the original paper, DP02, the total photon flux of the integrated SO band was found to be 2.3 x $10^{27}$ photons/sec, compared to 2.4 x $10^{27}$ photons/sec using the above method. 
The September 1999 data presented below has been re-calibrated using this slit-loss correction method and are shown for comparison with DP02. 

Finally, as the distance from Earth to Io varies from 4-6AU, for simple comparison of the fluxes from Io, all nights have been adjusted to the Earth-Io distance(4.08AU) of the 1999 data from DP02. It should be noted that the data presented in \cite{depater07} have not been adjusted to this distance hence the apparent discrepancy between the scales of the November 2002 AO disk integrated data in these two papers. The results are later converted to total emitted intensity (distance independent, see Section \ref{sub:totflux}) for ease of comparison with any future observations.

\section{Analysis}

\subsection{Continuum Analysis\label{sub:contcal}}

To analyze the emission band we subtracted a single temperature blackbody fit to the continuum of the surrounding spectra from 1.63-1.67$\mu m$ and 1.75-1.79$\mu m$, wavelength regions least affected by telluric lines and Io's SO emission band. In reality of course this blackbody emission is the superposition  of many different temperature hotspots, any of which can vary over multiple observations. There is no reason to believe a direct correlation between total blackbody emission and SO band strength exists, as the SO may be a product of only some types of eruptions \citep{zolotov98a}. 

Measurements of Io's continuum are complicated by the fact that by their nature all eclipse measurements of Io are taken when the satellite is very close to Jupiter (20"-40") and thus Io's spectra can be contaminated to varying degrees by Jupiter's glare. We therefore compare our results with a theoretical estimate of the variability of the blackbody
spectrum in the wavelength range of the N6 filter. To do this we used the temperatures and surface areas calculated in \cite{depater04} as the basis for a 19 hot spot model of Io. Although
these values were calculated in the 2.5-3$\mu m$ range , where we
are not as sensitive to high temperature hotspots, they are a sufficient
starting point for this estimate. We used a $10^{5}$ iteration monte
carlo simulation where the temperatures varied with gaussian fluctuations,
with a 1$\sigma$ value of 20\% of the starting temperature. For each distribution
of temperatures the photon flux can simply be estimated from:

\[
\Psi=\sum_{i=1}^{N}\frac{\lambda}{hc}B_{i}(\lambda,A_{i},T_{i})\cos(\phi_{i})\cos(\theta_{i})\frac{1}{d_{io}^{2}}\]

\begin{figure}[h]
\includegraphics[bb= 30 10 200 200]{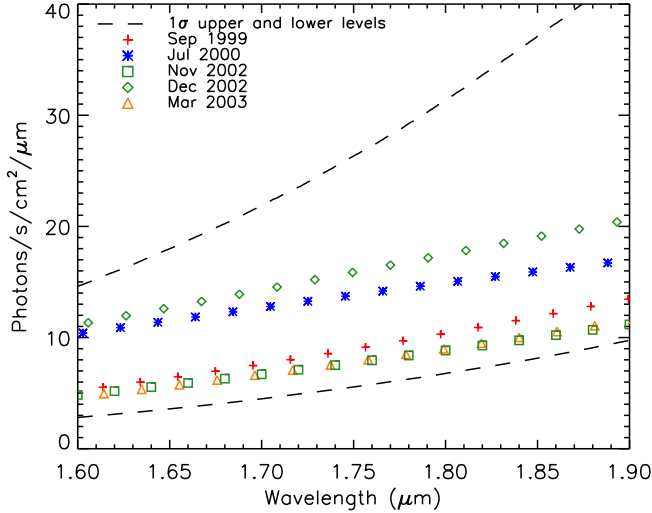}
\caption{Above shows the 1 sigma limits of  a Monte Carlo model showing expected flux with hotspot temperatures varying by 20\%, over-plotted are the fitted background continua from each night.}
\label{fig:monte20}
\end{figure}

Where $B_{i}$ is the emitted flux of the $i^{th}$hotspot which depends on the temperatures and areas from \cite{depater04}. $\theta_{i}$ and $\phi_{i}$ are the longitude and latitude of the $i^{th}$ hotspot w.r.t. the central meridian longitude and sub-earth latitude, and $d_{io}$ is the geocentric distance to Io (4.08AU). The cosine factors include the effects of foreshortening for each hotspot
using the positions determined in \cite{depater04} and
the typical sub-earth point for eclipse observations of Io in ingress ($\ensuremath{\sim}$347 degrees longitude). 
Fig. \ref{fig:monte20} shows the 1 $\sigma$ upper and lower bounds of the flux at each wavelength for 
$10^{5}$ iterations. Over-plotted are the fitted continua from the data taken in 1999-2003. The observed flux densities all fall between the 1 $\sigma$ limits of the Monte Carlo model, as expected. 

We also modeled a system in which all of the hotspots are stable except for one. We examined the effect of an outburst by varying Loki's temperature by up to 50\%, and saw a very similar profile to that shown in fig \ref{fig:monte20}.  This implies that a single, large hotspot can dominate the thermal background (\citealp{marchis02}, \citealp{macintosh03}, \citealp{depater07}), adding more uncertainty to the background measurements. We also investigated the effects of varying
the hotspot size by 10\% at constant temperature and varying the sub-earth
point over the hour duration of Io's eclipse and found these to be
of minor influence compared to the variations caused by temperature
fluctuations. 

\begin{figure*}[!ht]
\includegraphics[bb= 30 10 600 200]{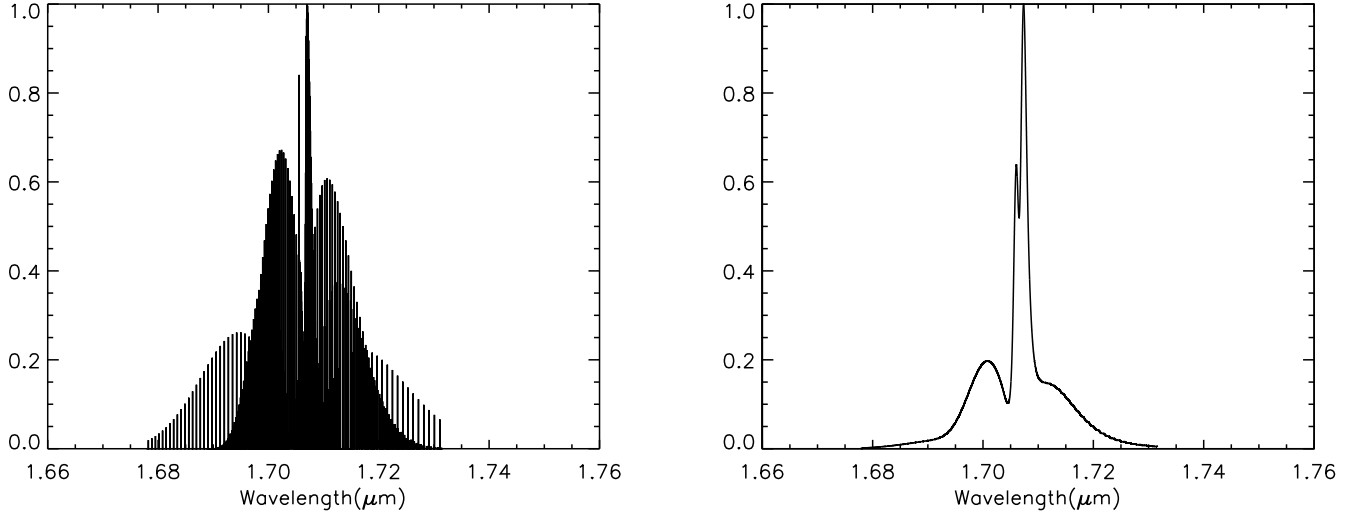}
\caption{\label{fig:labspec}Raw lab spectra showing the various transition branches of the SO 1.7$\mu$m band (left) and convolved with NIRSPEC resolution (right)}
\end{figure*}

\subsection{Rotational Temperature}\label{sub:linetemp}

As reported in DP02, the emission band detected
at 1.707$\mu$m is the SO forbidden $a^{1}\Delta\rightarrow X^{3}\Sigma^{-}$
ro-vibronic transition. The high resolution spectrum of this transition at
322K from \cite{setzer99} is shown 
in Fig. \ref{fig:labspec}(a). In panel (b) we show the same band after convolution with a gaussian
of full width at half maximum of 8.5\AA, which very closely approximates
the line spread function for NIRSPEC. 

\begin{figure*}[ph!]
\includegraphics[bb= 20 0 400 600]{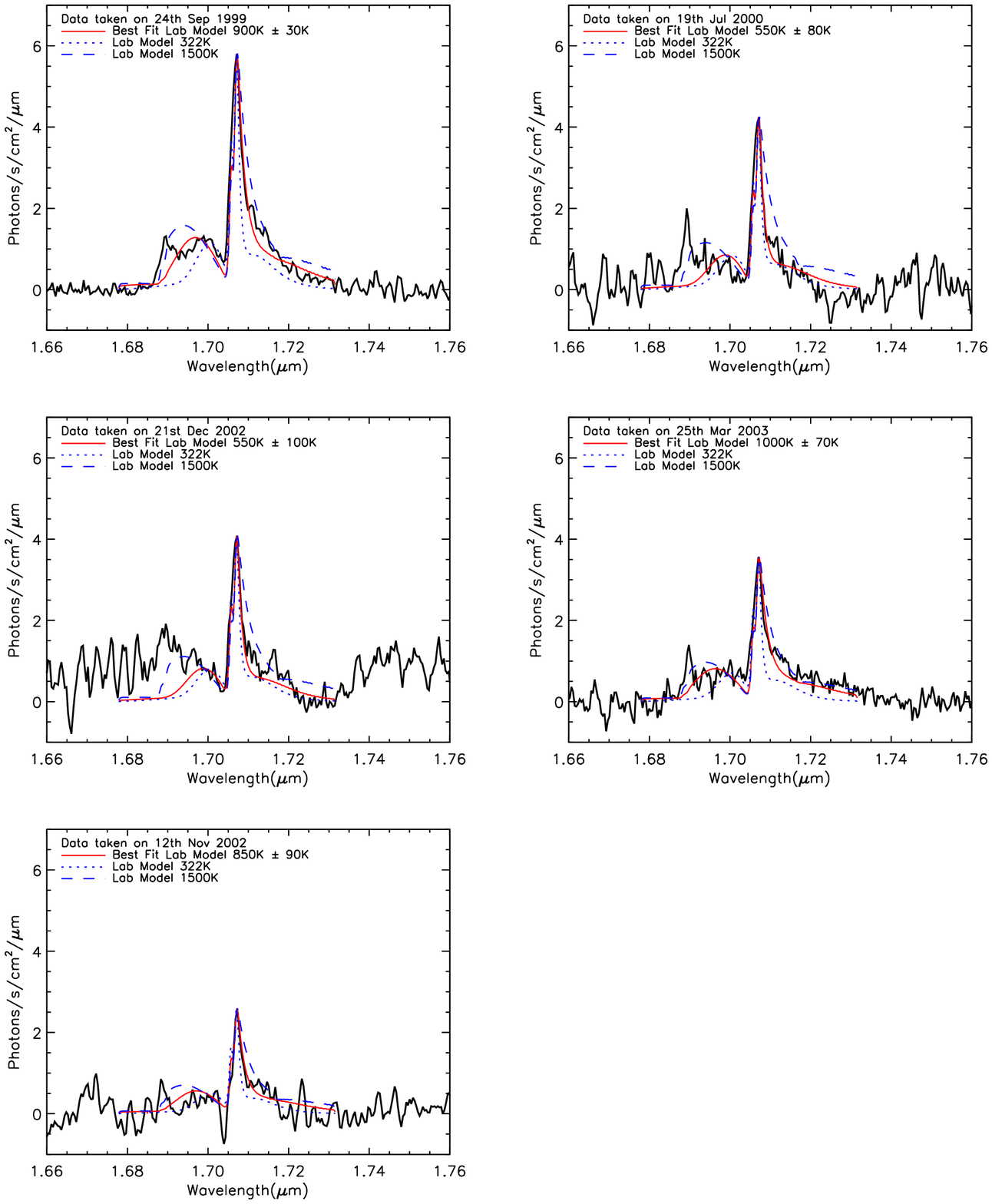}
\caption{\label{fig:allnight}SO band emission for all nights. Over-plotted are the best fit rotational temperature models using chi-squared minimization. The errors quoted in the legend the 1 sigma uncertainties.   Also shown are example line profiles at 322K and 1500K for comparison.}
\end{figure*}

To investigate the emission band's dependence on rotational temperature, we use DP02's model,
with wavenumbers from \cite{setzer99} and the rotational line-strength factors for the equivalent $a^{1}\Delta\rightarrow X^{3}\Sigma^{-}$ ro-vibronic transition of O$_{2}$ ($\sigma=\rho=0$) 
from \cite{bellary87}. The individual line strengths are calculated for a given rotational temperature and then the profile is convolved to NIRSPEC's resolution. We constructed a library of model spectra from 400K to 1500K at 50K intervals which we use to estimate the dominant rotational temperature of the emitting gas.  In each case we use chi-squared minimization to obtain the best fit to the data, and from this we were able to estimate the errors shown in Fig. \ref{fig:allnight}. We used data in the range of 1.70 to 1.72 $\mu$m to fit the line, as this contains the region of highest signal to noise and largest variation with temperature. The results, shown superimposed in Fig. \ref{fig:allnight}, give rotational temperatures between 550K and 1000K, thus supporting the volcanic origin (DP02, \citealp{zolotov98a}) of the SO gas. The errors are of order 30-100K justifying our choice of 50K increments for our model spectra. 

At $\ensuremath{\sim}$1.69$\mu$m there is also an unidentified peak which, as mentioned in DP02, corresponds to the $^{S}R$ branch of the SO band; however, it is much weaker in the laboratory spectra, possibly implying some non-thermal distribution of excited SO states. While we find this intriguing, analysis of distribution of states of SO gas in non-LTE conditions is beyond the scope of this paper.

\subsection{Total Photon Flux}\label{sub:totflux}

The total observed photon flux can be calculated by integrating over the observed emission band (from 1.685-1.725 $\mu$m), giving us values of $F_{obs}$ = 0.051, 0.031, 0.022, 0.035, and 0.017 photons/s/cm$^{2}$ for the 1999,2000, 2002, 2003 and 2002 AO nights respectively. For easier comparison all data have been adjusted to an observer to Io (geocentric) distance of 4.08AU. We can convert these to the total number of photons emitted by SO if we assume isotropic emission of the excited atoms in a plume, $N_{pe}$ = $F_{obs} 4\pi d_{io}^{2}$, where $d_{io}$ is the geocentric distance to Io, giving 2.4 x $10^{27}$, 1.5 x $10^{27}$, 1.0 x $10^{27}$, 1.6 x $10^{27}$ and 0.8 x $10^{27}$ photons/sec respectively (see Tab. \ref{tab:flux}). Note that these values are independent of surface emitting area or distance to the observer. Various excitation mechanisms for SO are discussed in DP02, in an attempt to convert these values to a observed column density. However, ultimately more information is needed about the local physics. Simple thermodynamic equilibrium calculations are rendered invalid by the rarefied nature of Io's atmosphere, where low collision rates would make it difficult to ensure a Boltzmann population of states.  Note the observed peak at 1.69$\mu$m, which may also indicate a non-equilibrium distribution of states. 

\begin{deluxetable}{ccccc}
\tablecaption{\label{tab:flux}Integrated band emission of SO's $a^{1}\Delta\rightarrow X^{3}\Sigma^{-}$ transition from 1999-2003}
\tablehead{\colhead{Date} & \colhead{SO emission} &\colhead{Rotational Temp.} &\colhead{Loki Measurement Date\tablenotemark{b}} & \colhead{Loki Flux\tablenotemark{b} }\\ \colhead{} & \colhead{(x$10^{27}$ Photons/sec)} & \colhead{(K)} & \colhead{} & \colhead{(GW/$\mu$m/str)}} 

\startdata
9/24/1999 & 2.4 $\pm$ 0.7 & 900 $\pm$ 30 & 9/15/1999 - 10/10/1999 & 22 - 60 \\
7/19/2000 & 1.5 $\pm$ 0.5& 550 $\pm$ 80 &  --- & --- \\
11/12/2002 & 0.8 $\pm$ 0.2\tablenotemark{a} & 850 $\pm$ 90 &  10/27/2002 & 6 \\
12/21/2002 & 1.0  $\pm$ 0.3& 550 $\pm$ 100 & 10/27/2002 - 1/13/2003 & 6 - 29 \\
3/25/2003 & 1.6  $\pm$ 0.5& 1000 $\pm$ 70 &  3/25/2003 & 30 \\
\enddata
\tablenotetext{a}{Disk averaged AO data}
\tablenotetext{b}{Data provided by Julie Rathbun (private communication). In each case we have shown the Loki observations  that are closest in time to our SO data. No 3.5$\mu$m measurements of Loki were taken within 4 months on either side of the 19th July 2000 observations. }
\end{deluxetable}

\section{Discussion}

The study of this ro-vibronic transition gives us a valuable tool to investigate SO, a species thought to originate from volcanic out-gassing (DP02). The exact composition and physical conditions of Io's atmosphere are still poorly constrained and while it is agreed that both volcanic processes and sublimation contribute to the atmosphere, the details of it's minor components are unclear. In this study we fit rotational temperatures ranging from 550K-1000K to the emission band of SO gas. These high rotational temperatures suggest SO to be of a volcanic origin. A volcanic or high pressure vent origin is plausible, because interconversion of rotational and translational energy only requires tens of collisions in comparison to $\sim 10^{8}$ collisions for interconversion of electronic ($a^1$) and translational energy in the O$_{2}$ molecule \citep{bood83}, so one would not expect the SO ($a^1 \Delta$) electronic state to be in thermodynamic equilibrium with SO's rotational levels in the tenuous plume atmospheres on Io.

An interesting parameter which remains unconstrained is the total SO column abundance, which could be determined if the ro-vibronic band $B^{1}\Sigma\rightarrow X^{3}\Sigma^{-}$ of SO  at 0.97 $\mu$m would be measured simultaneously with the 1.707$\mu$m band \citep{setzer99}. During the March 2003 run we therefore attempted to take spectra with NIRSPEC1, a filter which covers the range 0.947-1.121 $\mu$m. Unfortunately we could not detect the 0.97 $\mu$m transition, which is weak and dominated by the  strong glare of reflected sunlight from Jupiter at these shorter wavelengths. The non-uniformity of the contribution from Jupiter in our slit also prevents estimates of an upper limit on the ratio of the 0.97$\mu$m and 1.7$\mu$m band. 

In July 2000, we also observed with NIRSPEC in high resolution echelle spectroscopy mode (R=25000), in an attempt to resolve individual transitions, but the 1.7 $\mu$m band was not detectable. We conclude that a combination of limited observing time during Io's eclipse, the higher dispersion of our signal and a low band-to-noise ratio as seen in the low resolution data of the same time period, result in a signal to noise too low for us to detect. 

We are able to add to the estimates of emitted flux from the $a^{1}\Delta\rightarrow X^{3}\Sigma^{-}$ band, previously reported in DP02. The strongest emission is found in the 1999 data, which also happens to be the observation during which Loki was brightest, as shown in Fig. 1 of \cite{rathbun06}. In Table \ref{tab:flux} we summarize the SO emission, rotational temperature and the 3.5$\mu$m flux measured from Loki (J. Rathbun, personal communication). The highest photon flux is found in the 1999 data, which also happens to be the observation when Loki was brightest. Although naively one might expect the high photon fluxes to also correspond to high temperatures, no such correlation is seen. The temperature in 1999 is high, 900$\pm$30K, but a similar high temperature is seen in Nov 2002 when the photon flux was lowest. This may be caused by the fact that we may observe several active volcanic centers simultaneously, and our non-AO observations cannot resolve individual volcanoes. In addition, usually only a few of the hot spots are associated with plumes and the temperature in these plumes can vary from volcano to volcano. We see the average effect of all of these. 

With so few data points it is difficult to draw any solid conclusions. However, it should be noted that the highest band emission occured in late 1999, concurrent with a dramatic rise in Loki's observed 3.5$\mu$m intensity \citep{rathbun06}. Also the lowest observed SO band emission of 0.8 x 10$^{27}$ photons/sec occurs in late 2002 which corresponds to a time of very low Loki flux. 

Accurate measurements of the background IR continuum would allow us to determine the behavior of the SO emission in the context of the broader thermal IR emission of Io. However further study of the raw spectra indicates that contamination from reflected and diffuse Jupiter light affect the background level and slope, particularly in the 2000 and 2002 data, due to observational factors such as slit orientation and proximity to Jupiter.  By their nature these observations occur when Io is very close to Jupiter (20-40"), and that distance is changing rapidly in time resulting in a varying level of contamination from frame to frame. It should be noted that this does not affect our measurements of the SO emission band.

\section{Conclusion}
Sulfur monoxide was detected in the near infrared at 1.707 $\mu$m in all of the low resolution NIRSPEC 6 data sets taken from 2000 to 2003. Analysis of the NIRSPEC 1 data taken on 25th March 2003, failed to detect the 0.97 $\mu$m band, \citep{setzer99}, due to contamination by reflected sunlight from Jupiter. High spectral resolution data taken in July 2000, also failed to resolve the 1.7$\mu$m band.

Comparison of the band strengths on each of the nights show some variation. Although the photometric calibration is rough there appears to be a genuine change in SO emission from the 1999 data (2.4 x $10^{27}$photons/sec) to subsequent nights ($\sim$1 x $10^{27}$photons/sec). This is mirrored by the measured flux of Loki over this time period  indicating that SO emission may be linked with Loki's activity, although further investigation at high spatial resolution and/or concurrent monitoring of Loki and disk integrated SO spectra is required to prove there is a definitive correlation. 

Io's spectrum, when the satellite is in eclipse, is determined by hotspots, and so we expect to see a large variation in band-to-continuum ratio of SO as we look at finer spatial scales on Io. Using AO techniques one can spatially resolve individual hotspots, and perhaps be able to confirm the confinement of the SO 1.707 um emission band to volcanic areas. Ultimately, we hope to determine the exact spatial distribution of SO in Io's atmosphere, to determine whether Io's atmosphere is supported predominantly by volcanic activity or sublimation of SO$_{2}$ frost.

\section{Acknowledgments}
We would like to extend our warmest thanks to Julie Rathbun, for providing the data on Loki's activity. We also thank Emmanuel Lellouch and John Pearl for their very constructive and helpful reviews. 
Support for this research was provided by NSF grant AST 0406275 and by the HCICT Program (NASA Grant No. NNG05GA25G). The data were obtained with the W.M. Keck Observatory, which is operated by the California Institute of Technology, the University of California, Berkeley and the National Aeronautics and Space Administration. The Observatory was made possible by the generous financial support of the W.M. Keck Foundation. 
The authors extend special thanks to those of Hawaiian ancestry on whose sacred mountain we are privileged to be guests. Without their generous hospitality, none of the observations presented would have been possible. 

\bibliographystyle{apalike}

\end{document}